\documentclass{article}
\usepackage{spconf,amsmath,amsfonts,graphicx,multirow}

\title{Controllable Multichannel Speech Dereverberation based on Deep Neural Networks}
\name{Ziteng Wang, Yueyue Na, Biao Tian, Qiang Fu}
\address{Alibaba Group, China}
\begin{document}
%
\maketitle
\begin{abstract}
Neural network based speech dereverberation has achieved promising results in recent studies. Nevertheless, many are focused on recovery of only the direct path sound and early reflections, which could be beneficial to speech perception, are discarded. The performance of a model trained to recover clean speech degrades when evaluated on early reverberation targets, and vice versa. This paper proposes a novel deep neural network based multichannel speech dereverberation algorithm, in which the dereverberation level is controllable. This is realized by adding a simple floating-point number as target controller of the model. Experiments are conducted using spatially distributed microphones, and the efficacy of the proposed algorithm is confirmed in various simulated conditions.
\end{abstract}
\section{Introduction}

\begin{figure}[htb]
	\centering
	\centerline{\includegraphics[width=\columnwidth]{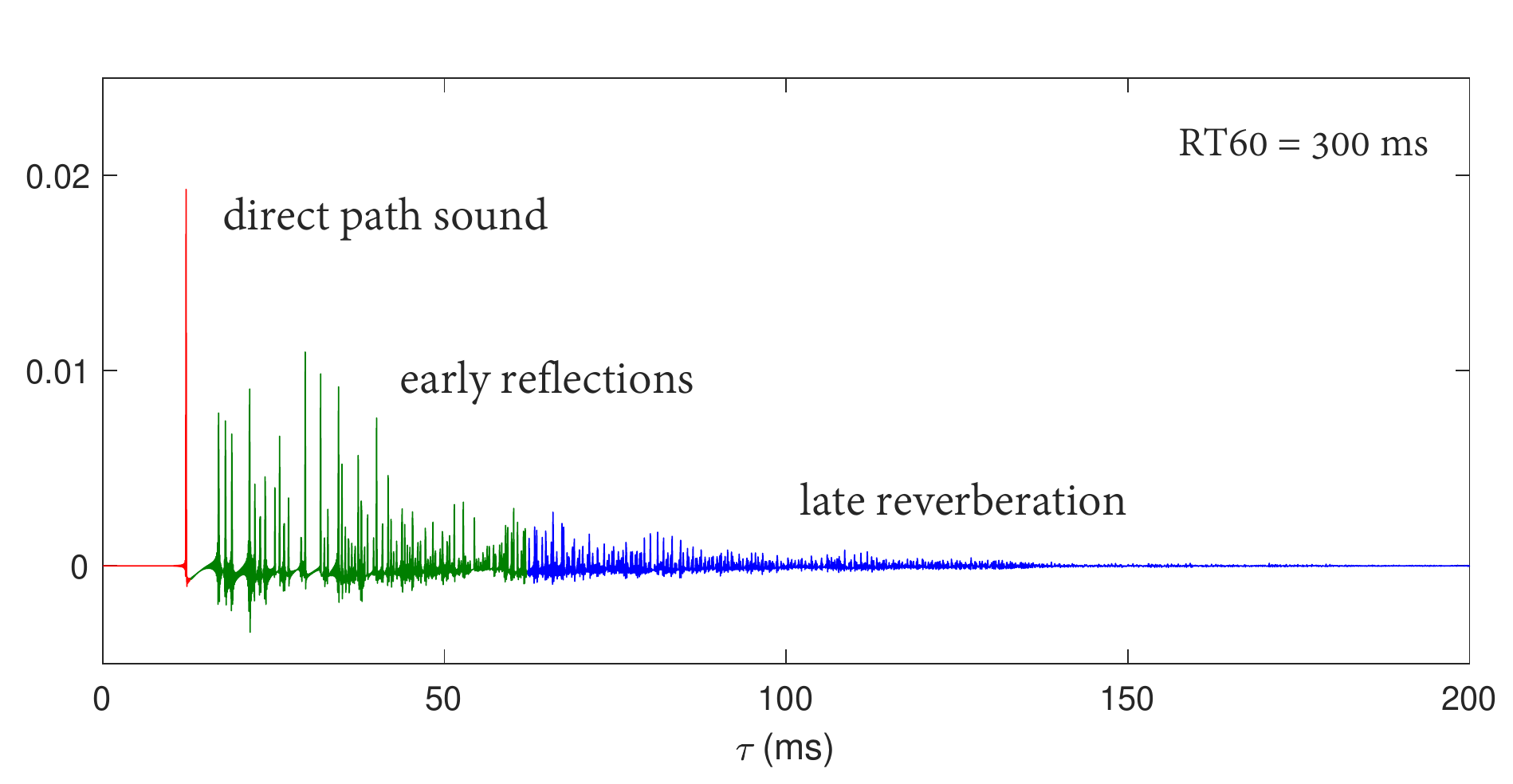}}
	\caption{Time samples of an example RIR.}
	\label{fig:rir}
\end{figure}

Reverberation is the cumulation of multiple reflections when the signal travels from  source to microphone in a reverberant room. Room reverberation can be characterized by the room impulse response (RIR) relating the source position and the microphone position. As shown in Fig.~\ref{fig:rir}, an example RIR consists of three successive parts: direct path sound, early reflections and late reverberation. While late reverberation generally needs to be suppressed, early reflections are harmless to automatic speech recognition~\cite{sivasankaran2017combined} and could be beneficial to speech perception in many cases~\cite{roman2013speech}.

One classical speech dereverberation algorithm is based on multichannel linear prediction (MCLP), such as the generalized weighted prediction error (GWPE)~\cite{yoshioka2012generalization} algorithm. GWPE aims to minimize the degree of temporal correlation in the output signals, based on the observation that non-reverberant speech is almost uncorrelated in time in the subband domain. It is designed to shorten the impulse responses rather than to cancel them, and therefore only late reverberation is suppressed. The algorithm has proved its efficacy of improving automatic speech recognition accuracy in the REVERB Challenge~\cite{kinoshita2016summary} and in commercial products such as the Google Home~\cite{caroselli2017adaptive}.

Neural network based speech dereverberation (NNDR) flourishes following the success of deep learning based speech enhancement~\cite{han2015learning,wu2016reverberation,williamson2017time,ernst2018speech}. The two tasks share many similarities. Han et al. \cite{han2015learning} proposed to learn a spectral mapping from reverberant to anechoic speech. Wu et al. \cite{wu2016reverberation} utilized an improved neural dereverberation system, in which frame shift and acoustic context size are tuned for different reverberation time. Some studies also consider performing dereverberation and denoising jointly or in a stage-wise manner~\cite{zhao2018two,li2021simultaneous}. Extending the single channel NNDR algorithms to the multichannel case, Kinoshita et al.~\cite{kinoshita2017neural} proposed to utilize neural network based spectrum estimation to construct linear inverse filters of WPE. Wang et al. \cite{wang2020multi,wang2020deep} explored spectral mapping based multichannel dereverberation in combination with beamformers.

The previous neural network based methods are, however, mostly focused on recovery of the direct path sound and early reflections are discarded. Some consider using separate models to recover the direct sound plus early reflections~\cite{williamson2017time,zhao2018late}. In this paper, we propose to use one model for controllable speech dereverberation, and thus save the usage of multiple models. By adding an intuitive floating-point number as a controller, the neural network model can be tuned to either remove or retain early reflections. Experiments are conducted using spatially distributed, i.e. ad hoc, microphones. 
The motivation is that late reverberation is closely related to the  acoustical characteristics of a room, and less dependent on the microphone positions.
It is further shown that the controller helps the model generalize to unseen training targets.


\section{Problem Formulation}

Given $C$ microphones and $D$ sources in a reverberant environment, the signal captured in the $c$th microphone is given by
\begin{equation}
	y_{c} (t) = \sum_{d=1}^{D} h_{c,d}(t) * s_d(t) + n_c(t)
\end{equation}
where $h_{c,d}$ is the RIR relating the microphone and source $s_d$, $*$ denotes linear convolution, and $n_c$ is the ambient noise. The desired signal, defined as the direct path sound with or without early reflections, is given by
\begin{equation}
x_{c} (t) = \sum_{d=1}^{D} \lfloor h_{c,d} \rfloor(t) * s_d(t)
\end{equation}
where $\lfloor h \rfloor$ is the correspondingly truncated RIR.

Without loss of generality, the following algorithms are developed in the frequency domain. The frequency representations of $y,x$ are denoted as $Y, X$ respectively. A neural network model is designed to estimate a predefined mask in each time-frequency bin, such as the complex ratio mask (CRM)
\begin{equation}
M = \frac{Y_r X_r + Y_i X_i}{Y_r^2 + Y_i^2} + j\frac{Y_r X_i - Y_i X_r}{Y_r^2 + Y_i^2}
\end{equation}
where $(\cdot)_r$ is the real part and $(\cdot)_i$ is the imaginary part of a complex variable, with $j$ the imaginary unit. The target signal can be recovered by applying inverse fast Fourier transform (IFFT) as
\begin{equation}
 \hat{x}_c(t) = \text{IFFT}({\bf M}_c \odot {\bf Y}_c)
\end{equation}
where ${\bf M}_c \in \mathbb{C}^{T \times K}$ is a collection of the estimated masks, ${\bf Y}_c$ is the complex signal spectrum, and $\odot$ denotes element-wise multiplication. $T$ refers to the signal length and $K$ is determined by the Nyquist frequency.

\section{The proposed algorithm}

The proposed neural network model for speech dereverberation accepts an input ${\bf F} \in \mathbb{R}^{T \times A}$ and produces an output ${\bar{\bf{M}}} \in \mathbb{R}^{T \times 2K}$, where $A$ refers to the feature vector size and ${\bar{\bf{M}}} = [{\bf M}_r, {\bf M}_i]$ is a concatenation of the real and imaginary parts of the target CRMs.

\begin{figure}[htb]
	\centering
	\centerline{\includegraphics[width=\columnwidth]{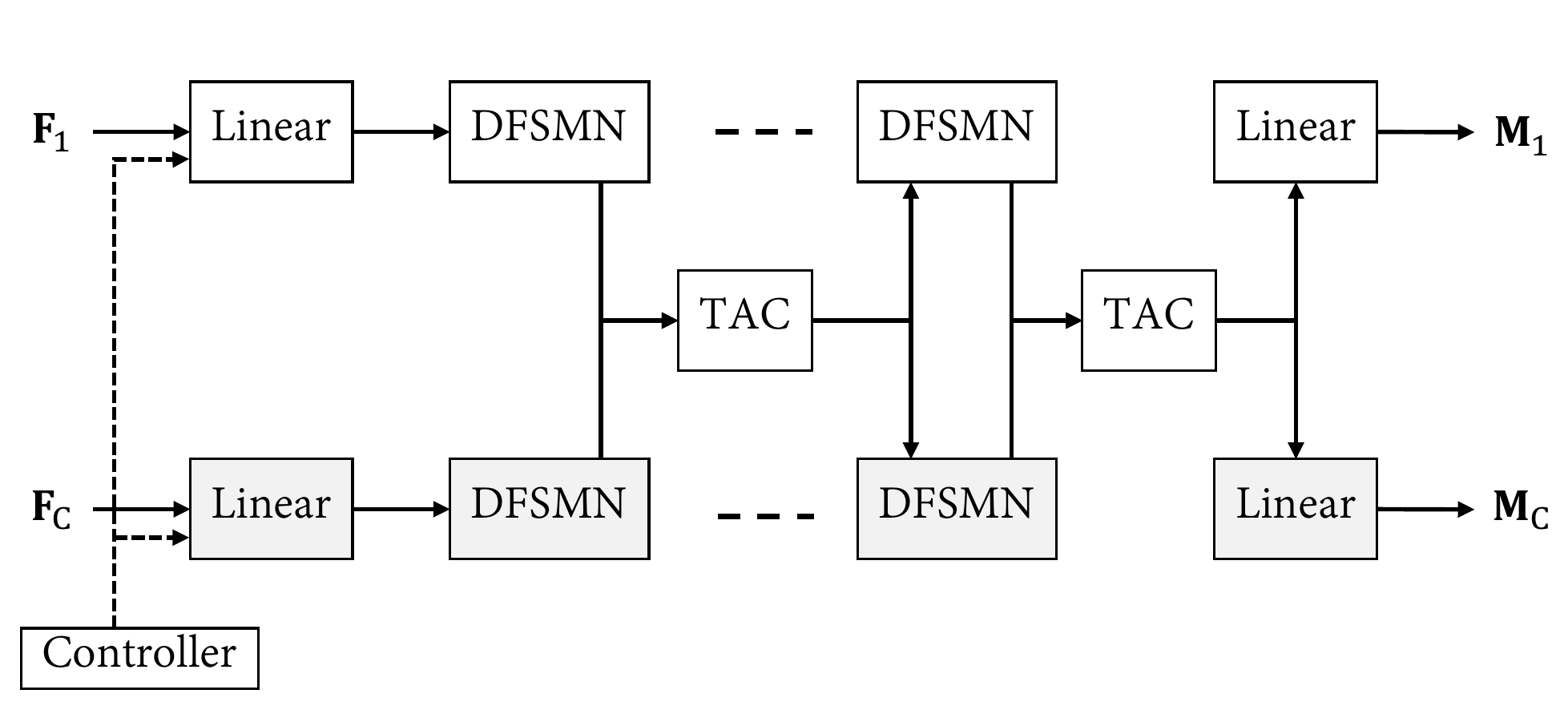}}
	\caption{The proposed MIMO-TAC model with a target-related controller. All channels share the same model parameters.}
	\label{fig:model}
\end{figure}

\subsection{MIMO-TAC}

We first consider a basic multiple input multiple output (MIMO) setup of the neural network model, which takes feature vectors from all channels as input and produces all target masks together. There are
\begin{align}
	\nonumber & {\bf H}_0 = \text{PReLU}(\text{Linear}([{\bf F}_1, {\bf F}_2, \cdots, {\bf F}_C])), \\
	\nonumber & {\bf H}_l = \text{DFSMN}({\bf H}_{l-1}), \quad l=1,...,L \\
	& [\bar{{\bf M}}_1, \bar{\bf{M}}_2, \cdots, \bar{\bf{M}}_C] = \text{Tanh}(\text{Linear}({\bf H}_L))
\end{align}
where Linear($\cdot$) denotes a linear transformation layer, and ${\bf H}_l$ denotes the output of the $l$th layer. The deep feed-forward sequential memory network (DFSMN) is chosen here, given its performance in several speech related tasks~\cite{zhang2018deep,wang2021weighted}. One DFSMN layer is expressed as
\begin{align}
	{\bf p}_{t,l} &= \text{Linear}(\text{ReLU}(\text{Linear}({\bf h}_{t,l-1}))), \\
	{\bf h}_{t,l} &= {\bf h}_{t,l-1} + {\bf p}_{t,l} + \sum_{\tau=0}^{\mathcal{T}}{\bf w}_{\tau,l} \odot {\bf p}_{t-\tau,l}
\end{align}
where ${\bf w}_{\tau,l}$ is the time-invariant memory parameter weighting the history output ${\bf p}_{t-\tau,l}$ and $\odot$ denotes element-wise multiplication.

Preliminary results show that the MIMO setup consistently outperforms a single input single output (SISO) setup, which performs speech dereverberation for each channel separately. Nevertheless, the dimensions of the model change as the number of microphones changes, which could limit its broad applicability, such as applying to the case with corrupted channels~\cite{yemini2021scene}. 
We adopt a transform average concatenate (TAC) method, originally proposed in \cite{luo2020end} for speech separation, to support arbitrary input channels and the new model is named as MIMO-TAC. One TAC layer is expressed as
\begin{align}
\nonumber {\bf t}_{t,l,c} &= \text{T-layer}({\bf h}_{t,l-1,c}), \\
\nonumber {\bf t}_{t,l} &= \frac{1}{C}\sum_{c=1}^{C} {\bf t}_{t,l,c}, \\
\nonumber {\bf a}_{t,l} &= \text{A-layer}({\bf t}_{t,l}), \\
{\bf h}_{t,l,c} &= {\bf h}_{t,l-1,c} + \text{C-layer}([{\bf t}_{t,l,c}, {\bf a}_{t,l}])
\end{align}
where the T-layer, A-layer and C-layer are all fully-connected layers followed by the PReLU activation function. We interleave the TAC layers with DFSMN layers as illustrated in Fig.~\ref{fig:model}. All the channels share the same model parameters.

\subsection{Controller}

To produce controllable outputs, a target-related controller is additionally used as input to the model. Here, one floating-point number is sufficient as will be shown in latter experiments. For each channel, the new feature vector is given by
\begin{equation}
	{\bar {\bf F}}_c = [{\bf F}_c, \text{f}]
\end{equation}
where $\text{f} \in [0, 1]$. For $\text{f}=0$, the training target is chosen as the direct path sound, and for $\text{f}=1$, the target is the direct sound plus early reflections.

\subsection{Loss function}

The neural network model is trained to minimize both a mask approximation loss $\mathcal{L}_{mask}$ and a signal approximation loss $\mathcal{L}_{sig}$ over all channels, where
\begin{align}
\nonumber \mathcal{L}_{mask} &= \sum_{c}|\hat{{\bf M}}_c - {\bf M}_c|^2 \\
\mathcal{L}_{sig} & = \sum_{t}\sum_{c}|\hat{x}_c(t) - {x}_c(t)|^2
\end{align}
The weighting of the two loss terms are tuned heuristically.

\section{Experiments}


\subsection{Setup}

The experiments are conducted in simulated rooms. 
RIRs are generated using an open-source implementation of the IMAGE method\footnote{https://github.com/ehabets/RIR-Generator}.
There are three room sets for training: \emph{small rooms, medium rooms} and \emph{large rooms}, where width and length of the small rooms are uniformly sampled from the range 1 m to 10 m, medium rooms uniformly sampled from 10 m to 30 m, and large rooms uniformly sampled from 30 m to 50 m. 
The absorption coefficients are sampled randomly from 0.2 to 0.8.
The absorption coefficients of the room walls are assumed the same, while those of the roof and floor are different. 
For each room set, 200 rooms are constructed and 100 RIRs are generated in each room based on one source position and $C=4$ microphone positions of an ad hoc array. The source speeches are collected from the WSJ0 corpus. 
In total, 55,000 RIRs are used for training and the other 5,000 are used as the development set.

The test set consists of 10 other random-sized rooms for each reverberation time in [0.3 s, 0.6 s, 0.9 s, 1.2 s]. 10 positions are sampled for each test room. Perceptual evaluation of speech quality (PESQ) and cepstrum distance (CD) are used as the performance metrics~\cite{kinoshita2016summary}. 

For Fourier transform, the frame length is 40 ms with half overlap, and there is $K=321$ with 16k sampling frequency. $A=80$ filter banks are used as input feature for each channel.
In the MIMO-TAC model, there are 9 DFSMN and 9 TAC layers each with 256 hidden nodes, leading to 3.7M trainable parameters. The DFSMN layers have a left history order of $\mathcal{T}=20$ frames and no look-ahead. So this is a causal model. The Adam optimizer with an initial learning rate of 0.001 is adopted during model training and gradient clipping is applied with a threshold of 10.

\subsection{Results and Analysis}

In Table 1, average PESQ scores using the direct sound plus 50ms early reverberation as both the training target and scoring reference are reported. 
For comparison, a vanilla SISO model (4.0M parameters) and a MIMO model (4.1M parameters) are included.
As shown in the table, all models score better than the original mixtures.
The MIMO-TAC model consistently outperforms its MIMO alternative by an average PESQ improvement of 0.07.
In the last column, a MIMO-TAC model is also separately trained on data from a 4-channel fixed circular array with diameter of 20 cm and tested on the ad-hoc array data.
The model performs quite well in the mismatched test cases. 
This confirms that multiple observations, regardless of the microphone positions, could benefit the dereverberation process. 

\begin{table}[t]
	\begin{center}
		\caption{PESQ scores evaluated on the ad hoc array using the direct sound plus 50ms early reflections as both the training target and scoring reference. The best scores are highlighted in bold.}
		\label{tab1}
		\begin{tabular}{c|c|c|c|c|c}
			\hline
			RT60 & Orig & SISO & MIMO 
			& \begin{tabular}{@{}c@{}}MIMO \\ -TAC\end{tabular} 
			& \begin{tabular}{@{}c@{}}Fixed \\ Array\end{tabular} \\ \hline
			0.3 s    & 2.94 &3.40 &3.43 &\bf 3.49 &3.44\\ \hline
			0.6 s    & 2.37 &2.90 &2.95 &\bf 3.02 &2.97 \\ \hline
			0.9 s    & 2.10 &2.60 &2.67 &\bf 2.74 &2.69 \\ \hline
			1.2 s    & 1.89 &2.37 &2.46 &\bf 2.52 &2.47 \\ \hline
		\end{tabular}
	\end{center}
\end{table}

\begin{table}[t]
	\begin{center}
		\caption{PESQ scores of the MIMO-TAC models evaluated using the direct sound with (bottom half) or without (top half) 50ms early reflections as scoring reference.}
		\label{tab2}
		\begin{tabular}{c|c|c|c|c|c}
			\hline
		    RT60 & Orig 
		    & \begin{tabular}{@{}c@{}}Direct \\ sound\end{tabular} 
		    & \begin{tabular}{@{}c@{}}50ms \\ early\end{tabular} 
		    & $\text{f}=0$ & $\text{f}=1$ \\ \hline
			0.3 s &2.46 &\bf 3.09 &2.83 &3.07 &- \\ \hline
			0.6 s &1.96 &\bf 2.63 &2.43 &2.62 &- \\ \hline
			0.9 s &1.77 &\bf 2.40 &2.30 &\bf 2.40 &- \\ \hline
			1.2 s &1.61 &2.18 &2.13 &\bf 2.20 &- \\ \hline \hline
			0.3 s &2.94 &3.10 &\bf 3.49 & - &3.46 \\ \hline
			0.6 s &2.37 &2.66 &\bf 3.02 &- &2.99 \\ \hline
			0.9 s &2.10 &2.47 &\bf 2.74 &- &2.70 \\ \hline
			1.2 s &1.89 &2.30 &\bf 2.52 &- &2.48 \\ \hline
		\end{tabular}
	\end{center}
\end{table}

In Table 2, three MIMO-TAC models are compared: one trained using direct sound as target, one trained using direct sound plus 50 ms early reflections as target, and the proposed one with an target controller.
Both the first two models perform well in the  matched test cases, as seen in the bottom and top halves of the table, where the direct sound with or without 50 ms early reflections are respectively used as scoring references. 
The first model, however, cannot compete with the second one when evaluated on targets different from training, and vice versa. The proposed model with an controller could handle both test cases properly as seen in the last two columns. Setting the controller value to 0, the model is on par with the first one. And setting the controller value to 1, the model scores very close to the second one. It should be stated that the three models are trained on the same amount of data.

\begin{figure}[tb]
	\centering
	\centerline{\includegraphics[width=\columnwidth]{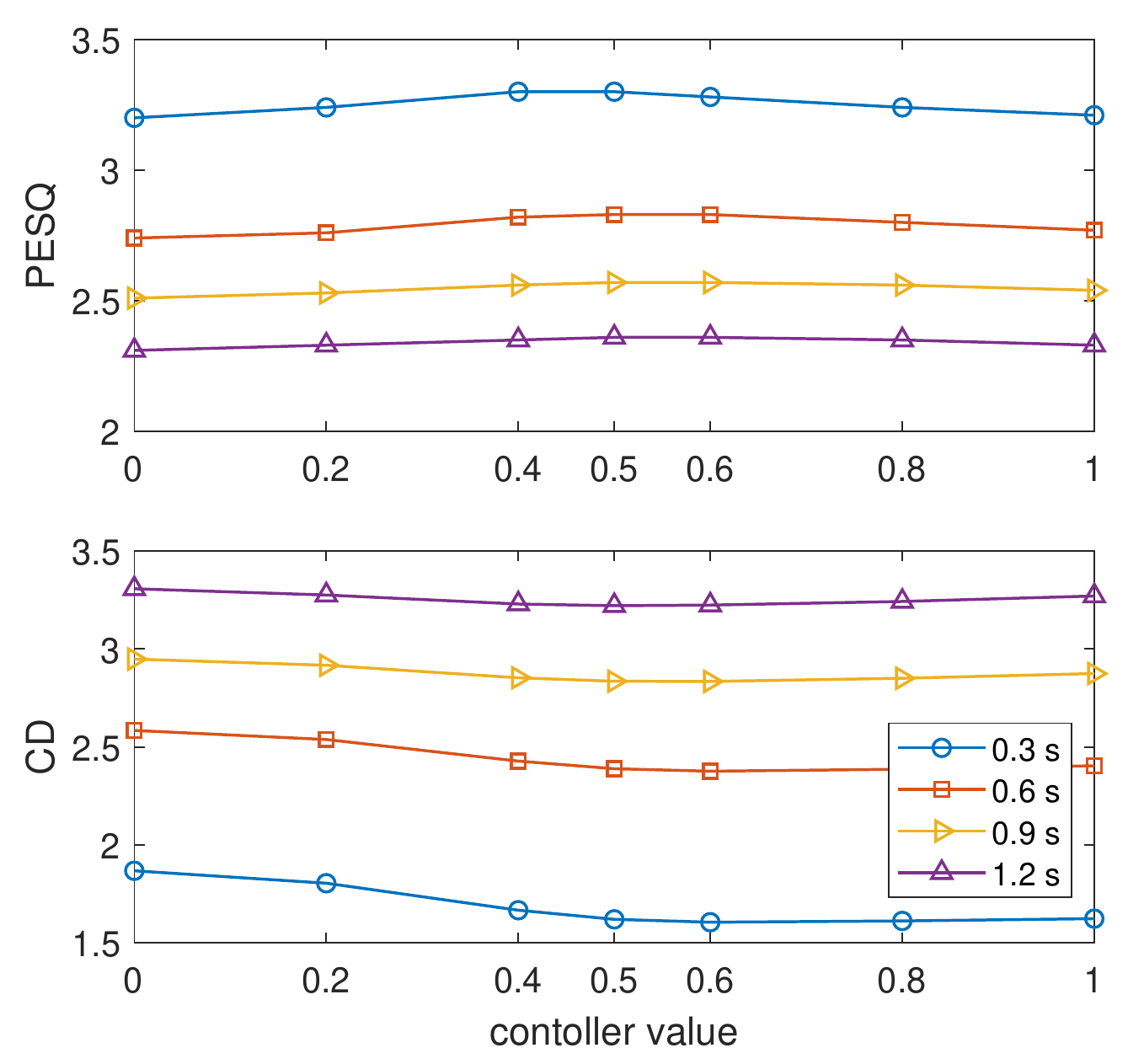}}
	\caption{PESQ/CD scores with respect to different controller values using the direct sound with 30 ms early reflections as scoring reference.}
	\label{fig:pesq}
\end{figure}

In Fig.~\ref{fig:pesq}, we further investigate how the controller would change the behavior of the proposed model by using a new target, the direct sound plus 30 ms early reflections, as scoring reference. As expected, the model generalize well to the unseen test case. The PESQ scores reach their peaks at a controller number of around 0.5. The peak score is 2.83 compared with 2.74 ($\text{f}=0$) and 2.77 ($\text{f}=1$) in reverberation time 0.6 s. Similar trends are observed in the CD scores, except that the best performance is achieved at a number of around 0.6. The best score is 2.83 compared with 2.94 ($\text{f}=0$) and 2.87 ($\text{f}=1$) in reverberation time 0.9 s. Although only 0/1 are used as inputs during training, the controller implicitly relates to the reverberation level in the output signals in a smooth and continuous way. 

\subsection{Discussions}

Controllable neural network models have been an emerging topic in target speech separation, such as Voicefilter~\cite{wang2019voicefilter} and SpEx~\cite{xu2020spex}, where target speeches are extracted by exploiting their corresponding speaker embedding vectors. Target speech extraction is much harder than the task discussed here, given that speaker vectors generally need to be representative and discriminative enough. Consequently, the speaker vectors are not continuous in their embedding space. The single-valued controller is special, because it does not require any prior information of the training targets. It would be appealing to try out the idea in other speech tasks, such as keyword spotting~\cite{liu2020metadata}.

\section{Conclusion}

This paper raises the question of controllable neural network based speech dereverberation and proposes a solution in which a floating-point number is introduced as target controller. 
The proposed MIMO-TAC model could be tuned to either remove or retain early reflections while suppressing late reverberation. 
The controller is shown to implicitly relate to the reverberation level in the output signal and help the model generalize well to unseen test conditions.


\bibliographystyle{IEEEbib}
\bibliography{refs}

\end{document}